\documentclass[11pt]{article}
\usepackage{bm, commath}
\usepackage{natbib}
\usepackage{caption}
\usepackage{graphicx}
\usepackage{subfig}
\usepackage{amsmath, amsfonts, amsthm}
\usepackage{float}
\usepackage{booktabs,siunitx}
\usepackage{url}
\usepackage{multirow}
\usepackage{amssymb}
\usepackage{blkarray}
\usepackage{bbm}
\usepackage{multicol}
\usepackage{authblk}

\usepackage{listings}
\usepackage{bbm}
\usepackage{textcomp}
\usepackage[margin=1in]{geometry}
\usepackage{authblk}
\usepackage[ruled]{algorithm2e}
\SetKwInput{KwParam}{Parameter}
\SetAlgoCaptionLayout{centerline}

\usepackage{sectsty}
\usepackage[onehalfspacing]{setspace}
\usepackage[utf8]{inputenc}
\usepackage{float}
\usepackage{tikz}
\usetikzlibrary{positioning,chains,fit,shapes,calc}

\newtheorem*{assumption*}{Assumption}

\usepackage{mathtools}

\usepackage{xcolor}

\makeatletter
\renewcommand{\algocf@captiontext}[2]{#1\algocf@typo. \AlCapFnt{}#2} 
\def\@algocf@capt@plain{top}
\renewcommand{\algocf@makecaption}[2]{%
  \addtolength{\hsize}{\algomargin}%
  \sbox\@tempboxa{\algocf@captiontext{#1}{#2}}%
  \ifdim\wd\@tempboxa >\hsize
  \hskip .5\algomargin%
  \parbox[t]{\hsize}{\algocf@captiontext{#1}{#2}}
  \else%
  \global\@minipagefalse%
  \hbox to\hsize{\box\@tempboxa}
  \fi%
  \addtolength{\hsize}{-\algomargin}%
}
\makeatother

\begin{document}

\sectionfont{\bfseries\large\sffamily}%

\subsectionfont{\bfseries\sffamily\normalsize}%




\title{Efficient algorithms for building representative matched pairs with enhanced generalizability}

\author[]{Bo Zhang \thanks{Bo Zhang is an assistant professor of biostatistics at the Fred Hutchinson Cancer Center. Email: {\tt bzhang3@fredhutch.org} } }

\affil[]{Vaccine and Infectious Disease Division, Fred Hutchinson Cancer Center, Seattle, WA, U.S.A.}

\date{}

\maketitle

\noindent
\textsf{{\bf Abstract}: Many recent efforts center on assessing the ability of real-world evidence (RWE) generated from non-randomized, observational data to produce results compatible with those from randomized controlled trials (RCTs). One noticeable endeavor is the RCT DUPLICATE initiative (Franklin et al., 2020, 2021). \textcolor{black}{To better reconcile findings from an observational study and an RCT, or two observational studies based on different databases, it is desirable to eliminate differences between study populations.} We outline an efficient, network-flow-based statistical matching algorithm that designs well-matched pairs from observational data that resemble the covariate distributions of a target population, for instance, the target-RCT-eligible population in the RCT DUPLICATE initiative studies or a generic population of scientific interest. We demonstrate the usefulness of the method by revisiting the inconsistency regarding a cardioprotective effect of the hormone replacement therapy (HRT) in the Women's Health Initiative (WHI) clinical trial and corresponding observational study. We found that the discrepancy between the trial and observational study persisted in a design that adjusted for study populations' cardiovascular risk profile, but seemed to disappear in a study design that further adjusted for the HRT initiation age and previous estrogen-plus-progestin use. The proposed method is integrated into the \textbf{R} package \textbf{match2C}.}%

\vspace{0.3 cm}

\section{Introduction}

\subsection{RCT DUPLICATE Initiative: Comparing Observational Studies to Randomized Controlled Trials (RCTs)}
\label{subsec: intro RCT duplicate}
In a recent high-profile study published in {\it Circulation}, the RCT DUPLICATE initiative (\citealp{franklin2020nonrandomized, franklin2021emulating}) designed $10$ observational studies using retrospective, non-randomized claims data, and compared their real-world-evidence-based (RWE-based) treatment effect estimates to those based on $10$ randomized controlled trials (RCTs) investigating very similar clinical questions. The RCT DUPLICATE initiative aims to build an empirical evidence base for real world data through large-scale replication of RCTs and understand to which clinical questions and by what analytic tools researchers could draw credible causal conclusions from retrospective, non-randomized data (e.g., electronic health records, administrative claims databases, and diseases registries). 


To enable a better comparison of effect estimates obtained from observational and RCT data, it is desirable to design an observational study whose treated and matched control groups are comparable to the RCT population in baseline covariates. In their design stage, \citet{franklin2021emulating} carefully emulated the RCT study population by applying the same inclusion and exclusion criteria to the observational data prior to statistical matching; however, tangible and potentially meaningful differences persist. \citet{franklin2021emulating} concluded:
\begin{quote}\small
    [I]nclusion and exclusion criteria from the trials could only be partially emulated, and even where fully emulated, the resulting distributions were at times meaningfully different between the RCT and RWE populations, possibly because of nonrepresentative participation in RCTs.
\end{quote}
Discrepancies in age, race/ethnicity, and important preexisting comorbid conditions between RCT and observational study populations are common in \citet{franklin2021emulating}'s emulation studies. For instance, the Saxagliptin and Cardiovascular Outcomes in Patients with Type 2 Diabetes Mellitus (SAVOR-TIMI 53) study (\citealp{scirica2013saxagliptin}), one of the ten trials \citet{franklin2021emulating} emulated, enrolled $33.1\%$ female and $37.8\%$ with history of myocardial infarction (MI); the observational study emulating the SAVOR-TIMI 53 trial, however, consisted of $46.8\%$ female and only $11.2\%$ with history of MI (\citealp[Table 1]{franklin2021emulating}). These differences persisted in the final matched samples constructed by \citet{franklin2021emulating}, and could partially explain the disagreement in effect estimates derived from RCTs and observational studies.

The ongoing RCT emulation study led by the RCT DUPLICATE initiative and many similar endeavors to better reconcile RCT and observational study findings, for instance, the Women’s Health Initiative study \citep{prentice2005combined,hernan2008observational}, or findings derived from different observational databases, motivate us to develop a transparent, efficient and easy-to-use algorithm that constructs homogeneous matched pairs that resemble a template (that is, a random sample from the target population) in key covariates.

There are two important applications for such an algorithm. First, in the context of target trial emulation using observational data \citep{franklin2021emulating}, the target population would be the RCT-eligible population, the template would consist of trial participants, and the algorithm could be used to construct treated and matched control groups that resemble the target RCT population. Second, the algorithm may be used to help assess inconsistencies often seen among observational studies based on different databases. For instance, in a recent study of the impact of intraoperative transesophageal echocardiography (TEE) on patients' clinical outcomes, \citet{metkus2021transesophageal} conducted a matched cohort study and found an effect smaller than that derived by \citet{mackay2021association}. \citeauthor{metkus2021transesophageal}'s \citeyearpar{metkus2021transesophageal} analysis was based on the Society of Thoracic Surgeons Adult Cardiac Surgery Database (STS ACSD) while \citeauthor{mackay2021association}'s \citeyearpar{mackay2021association} analysis was based on Medicare beneficiaries who are at higher peri-operative risk compared to the adult population in the STS ACSD. In this example, Medicare beneficiaries would be the target population, the template would consist of a random sample from Medicare claims data, and our algorithm could be used to construct matched cohorts from the STS ACSD that mimic Medicare beneficiaries in risk factors. This could enable researchers to estimate the effect of TEE for an important population (elderly Americans enrolled in Medicare) using STS ACSD, the world's premier clinical outcomes registry for adult cardiac surgery, and facilitate a more informed comparison between results derived from different databases.

\subsection{A Naive Method, An Existing Method, and A New Approach}
One naive strategy to create homogeneous matched pairs resembling a template would consist of two steps. Take the RCT DUPLICATE initiative as an example. In the first step, researchers could select a subset of treated participants from the observational data via matching on covariates collected by both the RCT and the observational study. In the second step, controls in the observational database are then matched to the treated participants selected by the first step. Each step involves only two groups and could be done by using standard statistical matching algorithms, for instance, network-flow-based algorithms built upon bipartite networks, like \citeauthor{silber2014template}'s \citeyearpar{silber2014template} template matching method, or mixed-integer-programming-based (MIP-based) approaches \citep{zubizarreta2012using, bennett2020building}.

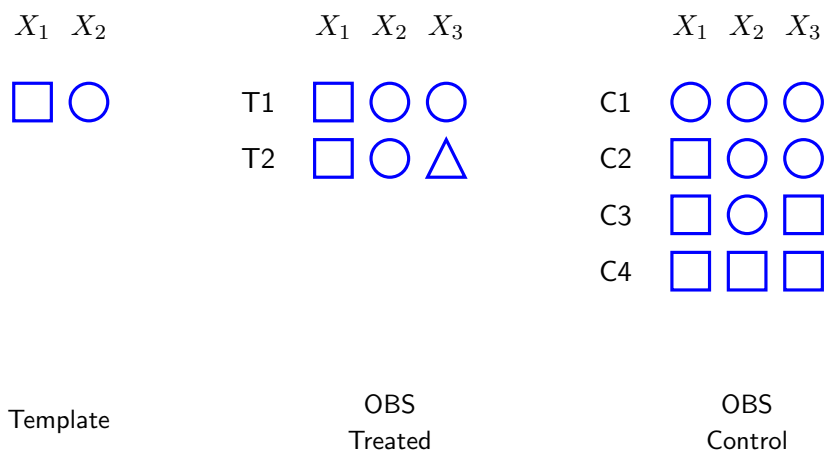
\begin{figure}[ht]
    \centering
   \begin{tikzpicture}

\node at (0.25, 1.25) {$X_1$};
\node at (1, 1.25) {$X_2$};
\draw[blue, very thick] (0,0) rectangle (0.5,0.5);
\draw[blue, very thick] (1,0.25) circle (0.25);
\node[text width=2cm, align = center] at (0.6, -4) {\small\textsf{Template}};

\node at (4.25, 1.25) {$X_1$};
\node at (5, 1.25) {$X_2$};
\node at (5.75, 1.25) {$X_3$};

\node at (3.25,0.25) {$\textsf{T1}$};
\node at (3.25,-0.5) {$\textsf{T2}$};
\draw[blue, very thick] (4,0) rectangle (4.5,0.5);
\draw[blue, very thick] (5,0.25) circle (0.25);
\draw[blue, very thick] (5.75,0.25) circle (0.25);

\draw[blue, very thick] (4,-0.75) rectangle (4.5,-0.25);
\draw[blue, very thick] (5,-0.5) circle (0.25);
\draw[blue, very thick] (5.5,-0.75) -- (5.75,-0.25) -- (6,-0.75) -- cycle;

\node[text width=2cm, align = center] at (5, -4) {\small\textsf{OBS \\ Treated}};

\node at (9, 1.25) {$X_1$};
\node at (9.75, 1.25) {$X_2$};
\node at (10.5, 1.25) {$X_3$};

\node at (8,0.25) {$\textsf{C1}$};
\node at (8,-0.5) {$\textsf{C2}$};
\node at (8,-1.25) {$\textsf{C3}$};
\node at (8,-2) {$\textsf{C4}$};

\draw[blue, very thick] (9,0.25) circle (0.25);
\draw[blue, very thick] (9.75,0.25) circle (0.25);
\draw[blue, very thick] (10.5,0.25) circle (0.25);

\draw[blue, very thick] (8.75,-0.75) rectangle (9.25,-0.25);
\draw[blue, very thick] (9.75,-0.5) circle (0.25);
\draw[blue, very thick] (10.5,-0.5) circle (0.25);

\draw[blue, very thick] (8.75,-1.5) rectangle (9.25,-1);
\draw[blue, very thick] (9.75,-1.25) circle (0.25);
\draw[blue, very thick] (10.25,-1.5) rectangle (10.75, -1);

\draw[blue, very thick] (8.75,-2.25) rectangle (9.25,-1.75);
\draw[blue, very thick] (9.5,-2.25) rectangle (10, -1.75);
\draw[blue, very thick] (10.25,-2.25) rectangle (10.75, -1.75);

\node[text width=2cm, align = center] at (9.75, -4) {\small\textsf{OBS \\ Control}};

\end{tikzpicture}
\caption{A toy example illustrating the limitation of a naive two-step approach. Suppose our goal is to create one treated-to-control pair. In the first step, treated units are matched to the template to obtain a candidate treated group. In this toy example, both $\textsf{T1}$ and $\textsf{T2}$ are good match to the template judging from two RCT covariates $X_1$ and $X_2$. However, in view of the reservoir of control units, $\textsf{T1}$ is a better pick as no control unit has an $X_3$ value similar to that of $\textsf{T2}$.}
\label{fig: illustrate naive method}
\end{figure}

This two-step approach suffers from a major drawback. \citet{franklin2021emulating} matched on more than $120$ covariates collected by the claims data to guard against unmeasured confounding in their emulation of the target SAVOR-TIMI 53 trial, although the trial reported only around $20$ baseline covariates. In the first step, there are many ways to design a smaller treated group similar to the RCT-eligible population in $20$ RCT covariates; however, it is difficult to determine which treated group designed from the first step should be used to form the final match in the second step. A selected treated group similar to the RCT-eligible population in $20$ RCT covariates could have poor overlap with controls in the observational database in the other more than $100$ covariates; see Figure \ref{fig: illustrate naive method} for a toy example illustrating this phenomenon. In many applications, resemblance of the matched cohort data to some target population is useful but not necessary; priority should be given to creating well-matched treated and control groups to first maximize a study's validity. The naive two-step approach precludes a principled trade-off between a matched cohort study's internal validity and its generalizability to a target population.

In principle, \citeauthor{bennett2020building}'s \citeyearpar{bennett2020building} MIP-based method could be adapted to designing a treated group and a matched control group that are well balanced in many covariates while resembling a template in key covariates by modifying constraints in their mathematical program. Although recent advancements in computing power have made it more practical to solve large-scale MIP problems with many complex constraints, the general MIP problems are still theoretically intractable or NP-hard. From a very practical perspective, MIP-based methods require installing a powerful commercial optimization routine (e.g., Gurobi or IBM CPLEX), which is proprietary and could be an obstacle to researchers. Other approaches to adjusting for sample selection bias include weighting and doubly-robust methods (see, e.g., \citealp{stuart2011use,Dahabreh:2019aa}).

Compared to MIP-based statistical matching methods, network-flow-based methods (see Supplemental Material A for a literature review) only require solving a polynomial solvable problem that is tractable both in theory and practice, and have proven successful in empirical comparative effectiveness research for decades (\citealp{rosenbaum2002observational,rosenbaum2010design, stuart2010matching, austin2011introduction,rassen2012one}). The primary goal of this article is to outline an efficient, network-flow-based algorithm that designs matched pairs from observational data with close resemblance to a target population. We demonstrate the usefulness of the proposed method by revisiting the Women's Health Initiative (WHI) study and exploring how our method facilitates different study designs and yields insight into the inconsistency between the WHI observational study and trial findings. We found that the discrepancy regarding a cardioprotective effect of the hormone replacement therapy persisted in a design that adjusted for the cardiovascular risk profile differences between the observational and trial data, but seemed to disappear in a design that further adjusted for the HRT initiation age and previous estrogen-plus-progestin use, resonating with similar findings in the seminal work by \citet{hernan2008observational}.

We conclude this section with an important caveat. Many reasons may contribute to the disagreement between effect estimates derived from an observational study and an RCT or from two observational studies. The synergy between effect heterogeneity and difference in covariate distributions is only one of the many possible reasons. Other important reasons include unmeasured confounding, noncompliance in the RCT, difference in the definition of treatment and/or clinical endpoints, among others. A matched cohort constructed from observational data, no matter how balanced its treated and matched control groups are, and how similar these two groups are to the target population, is \emph{not} necessarily free from unmeasured confounding bias and in our opinion, cannot replace an RCT. Nevertheless, addressing one critical and often conspicuous issue, that is, difference in covariate distributions among different studies, helps researchers focus on other issues and is a meaningful step towards reconciling different study outcomes and understanding the underlying mechanisms. 
\section{Methodology}
\label{sec: basic structure}
\subsection{Basic Network Structure: Vertices and Edges}
\label{subsec: basic structure v and e}
We describe our proposed method using the RCT emulation study described in Section \ref{subsec: intro RCT duplicate} for illustration purpose; the method can accommodate an arbitrary template from a target population other than the RCT-eligible population (e.g., the population of Medicare beneficiaries as discussed in the TEE and cardiac surgery example in Section \ref{subsec: intro RCT duplicate}).

\begin{figure}[ht]
\centering
\begin{tikzpicture}[thick, color = black,
  fsnode/.style={circle, fill=black, inner sep = 0pt, minimum size = 5pt},
  ssnode/.style={circle, fill=black, inner sep = 0pt, minimum size = 5pt},
  shorten >= 3pt,shorten <= 3pt
]


\begin{scope}[start chain=going below,node distance=7mm]
\foreach \i in {1,2,3}
  \node[fsnode,on chain] (r\i) [label=above left: {\small$\kappa_\i$} ] {};
\end{scope}

\begin{scope}[xshift=3cm,yshift=0cm,start chain=going below,node distance=7mm]
\foreach \i in {1,2,3,4}
  \node[ssnode,on chain] (t\i) [label=above right: {\small$\tau_\i$}] {};
\end{scope}

\begin{scope}[xshift=6cm,yshift=0cm,start chain=going below,node distance=7mm]
\foreach \i in {1,2,3,4}
  \node[ssnode,on chain] (tt\i) [label=above left: {\small$\overline\tau_\i$}] {};
\end{scope}

\begin{scope}[xshift=9cm,yshift=0cm,start chain=going below,node distance=7mm]
\foreach \i in {1,2,3,4,5,6}
  \node[ssnode,on chain] (c\i) [label=above right: {\small$\gamma_\i$}] {};
\end{scope}

\node [circle, fill = black, inner sep = 0pt, minimum size = 5pt, label=left: $\xi$] at (-2, -0.9) (source) {};

\node [circle, fill = black, inner sep = 0pt, minimum size = 5pt, label=right: $\overline\xi$ ] at (12, -2.2) (sink) {};



\foreach \i in {1,2,3} {
   \draw[color=gray] (source) -- (r\i);
   }

\foreach \i in {1,2,3} {
   \foreach \j in {1,2,3,4} {
   \draw[color=gray] (r\i) -- (t\j);
   }
 } 
 
\foreach \i in {1,2,3,4} {
   \draw[color=gray] (t\i) -- (tt\i);
}  
 
\foreach \i in {1,2,3,4} {
   \foreach \j in {1,2,3,4,5,6} {
   \draw[color=gray] (tt\i) -- (c\j);
   }
} 

\foreach \i in {1,2,3,4,5,6} {
   \draw[color=gray] (c\i) -- (sink);
}

\node[text width=2cm, align = center] at (0,1.5) {\small\textsf{RCT \\ Treated}};
\node[text width=2cm, align = center] at (3,1.5) {\small\textsf{OBS \\ Treated}};
\node[text width=2cm, align = center] at (6,1.5) {\small\textsf{OBS \\ Treated}};
\node[text width=2cm, align = center] at (9,1.5) {\small\textsf{OBS \\ Control}};

\end{tikzpicture}
\caption{A dense network for pair matching while emulating a target RCT population. Treated units from the target RCT population appear on the far left as node $\kappa_s$, $s = 1, 2, \dots, R$. Treated units from the observational database appear twice as $\tau_t$ and $\overline\tau_t$, $t = 1, 2, \dots, T$. Control units from the observational database appear on the far right as $\gamma_c$, $c = 1, 2, \dots, C$. There is a source $\xi$ and a sink $\overline{\xi}$. We have $R = 3$, $T = 4$, and $C = 6$ in this simple example.}
\label{fig: tripartite new scheme}
\end{figure}
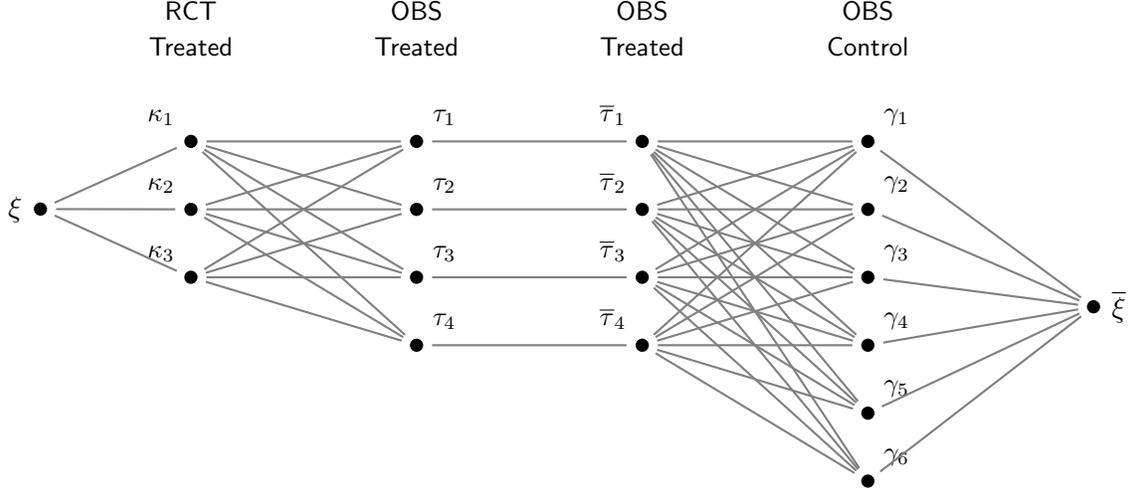

Figure \ref{fig: tripartite new scheme} depicts a basic version of the proposed network structure. There are $R$ treated units from the target RCT. These $R$ RCT units $\mathcal{K} = \{\kappa_1, \kappa_2, \dots, \kappa_R\}$ are represented by nodes labeled $\kappa_r$, $r = 1, 2, \dots, R$. There are $T \geq R$ OBS treated units and $C \geq T$ OBS control units from some administrative database. OBS treated units $\mathcal{T} = \{\tau_1, \tau_2, \dots, \tau_T\}$ are represented twice in the network, by nodes labeled $\tau_t$ and $\overline\tau_t$, $t = 1, 2, \dots, T$, and OBS control units $\mathcal{C} = \{\gamma_1, \gamma_2, \dots, \gamma_C\}$ are represented by nodes labeled $\gamma_c$, $c = 1, 2, \dots, C$. In addition to $R + 2 \times T + C$ nodes representing RCT and OBS study units, there is a source node $\xi$ and a sink node $\overline{\xi}$, so that the network consists of $|\mathcal{V}| = R + 2T + C + 2 = O(C)$ nodes in total:
\begin{equation}
    \mathcal{V} = \left\{\xi, \kappa_1, \dots, \kappa_R, \tau_1, \dots, \tau_T, \overline\tau_1, \dots, \overline\tau_T, \gamma_1, \dots, \gamma_C, \overline\xi\right\}.
\end{equation}

An ordered pair of vertices is referred to as an edge in the network. The basic structure in Figure \ref{fig: tripartite new scheme} consists of the following edges:
\begin{equation}
    \mathcal{E} = \left\{(\xi, \kappa_r), (\kappa_r, \tau_t), (\tau_t, \overline\tau_t), (\overline\tau_t, \gamma_c), (\gamma_c, \overline\xi),~r = 1, \dots, R, ~t = 1, \dots, T,~c = 1, \dots, C \right\}.
\end{equation}
There are a total of $|\mathcal{E}| = R + R \times T + T + T \times C + C = O(C^2)$ edges assuming $C$ is a constant multiple of $T$. 

\subsection{Basic Network Structure: Capacity and Matched Samples}
\label{subsec: basic structure: c and c}
To transform a statistical matching problem into an appropriate network-flow optimization problem, one needs to carefully design the cost and capacity of each edge. Fix an integer $k$ such that $1 \leq k \leq \lfloor T/R \rfloor$, and consider constructing $N = kR$ treated-to-control matched pairs from the observational database. For instance, the SAVOR-TIMI 53 trial consists of $R = 8,280$ units assigned saxaglipitin, the intervention under evaluation, and three retrospective databases available for emulation (Optum, MarketScan, and Medicare) consist of $T = 91,082$ subjects exposed to saxaglipitin according to \citet{franklin2021emulating}'s study protocol registered at ClinicalTrials.gov (identifier NCT03936023). Researcher could in principle choose any integer $k$ between $1$ and $\lfloor 91,082/8,280 \rfloor = 11$ in this example.

Let $\text{cap}(e)\geq 0$ denote the capacity of an edge $e \in \mathcal{E}$. In the basic network structure depicted in Figure \ref{fig: tripartite new scheme}, $\text{cap}\{(\xi, \kappa_r)\} = k$ for $r = 1, \dots, R$, and all other edges have capacity $1$. In Figure \ref{fig: tripartite new scheme}, the source $\xi$ supplies $kR$ units of flow, the sink $\overline\xi$ absorbs $kR$ units of flow, while all other nodes preserve the flow by simply passing them along (\citealp{ahuja1988network,bertsekas1991linear,rosenbaum1989optimal}). A feasible flow $f(\cdot)$ of the proposed network is formally defined as a mapping from the set of edges $\mathcal{E}$ to $\{0, 1, 2, \dots, k\}$ such that (i) all capacity constraints are respected, i.e., $0 \leq f(e) \leq \text{cap}(e),~e\in\mathcal{E}$, (ii) $kR$ units of flow are supplied at $\xi$ and absorbed at $\overline\xi$, i.e., $\sum_{r = 1}^R f\{(\xi, \kappa_r)\} = kR$ and $\sum_{c = 1}^C f\{(\gamma_c, \overline\xi)\} = kR$, and (iii) the flow is preserved at all nodes other than $\xi$ and $\overline\xi$, i.e., $\sum_{(a,b)\in\mathcal{E}\backslash\{\xi, \overline\xi\}} f\{(a,b)\} = \sum_{(b,c)\in\mathcal{E}\backslash\{\xi, \overline\xi\}} f\{(b,c)\}$ for all $b \in \mathcal{V}\backslash\{\xi, \overline\xi\}$.

\begin{figure}[ht]
\centering
\begin{tikzpicture}[thick, color = black,
  fsnode/.style={circle, fill=black, inner sep = 0pt, minimum size = 5pt},
  ssnode/.style={circle, fill=black, inner sep = 0pt, minimum size = 5pt},
  shorten >= 3pt,shorten <= 3pt
]


\begin{scope}[start chain=going below,node distance=7mm]
\foreach \i in {1,2,3}
  \node[fsnode,on chain] (r\i) [label=above left: {\small$\kappa_\i$} ] {};
\end{scope}

\begin{scope}[xshift=3cm,yshift=0cm,start chain=going below,node distance=7mm]
\foreach \i in {1,2,3,4}
  \node[ssnode,on chain] (t\i) [label=above right: {\small$\tau_\i$}] {};
\end{scope}

\begin{scope}[xshift=6cm,yshift=0cm,start chain=going below,node distance=7mm]
\foreach \i in {1,2,3,4}
  \node[ssnode,on chain] (tt\i) [label=above left: {\small$\overline\tau_\i$}] {};
\end{scope}

\begin{scope}[xshift=9cm,yshift=0cm,start chain=going below,node distance=7mm]
\foreach \i in {1,2,3,4,5,6}
  \node[ssnode,on chain] (c\i) [label=above right: {\small$\gamma_\i$}] {};
\end{scope}

\node [circle, fill = black, inner sep = 0pt, minimum size = 5pt, label=left: $\xi$] at (-2, -0.9) (source) {};

\node [circle, fill = black, inner sep = 0pt, minimum size = 5pt, label=right: $\overline\xi$ ] at (12, -2.2) (sink) {};



\foreach \i in {1,2,3} {
   \draw[color = black, line width = 0.5mm] (source) -- (r\i);
   }

\foreach \i in {1,2,3} {
   \foreach \j in {1,2,3,4} {
   \draw[color=gray!50] (r\i) -- (t\j);
   }
 } 
 
\draw[color = black, line width = 0.5mm] (r1) -- (t2);
\draw[color = black, line width = 0.5mm] (r2) -- (t3);
\draw[color = black, line width = 0.5mm] (r3) -- (t1);

\foreach \i in {1,2,3,4} {
   \draw[color=gray!50] (t\i) -- (tt\i);
}  

\draw[color = black, line width = 0.5mm] (tt2) -- (t2);
\draw[color = black, line width = 0.5mm] (tt3) -- (t3);
\draw[color = black, line width = 0.5mm] (tt1) -- (t1);

\foreach \i in {1,2,3,4} {
   \foreach \j in {1,2,3,4,5,6} {
   \draw[color=gray!50] (tt\i) -- (c\j);
   }
} 

\foreach \i in {1,2,3,4,5,6} {
   \draw[color=gray!50] (c\i) -- (sink);
}

\draw[color = black, line width = 0.5mm] (tt2) -- (c5);
\draw[color = black, line width = 0.5mm] (tt3) -- (c4);
\draw[color = black, line width = 0.5mm] (tt1) -- (c1);

\draw[color = black, line width = 0.5mm] (c1) -- (sink);
\draw[color = black, line width = 0.5mm] (c4) -- (sink);
\draw[color = black, line width = 0.5mm] (c5) -- (sink);

\node[text width=2cm, align = center] at (0,1.5) {\small\textsf{RCT \\ Treated}};
\node[text width=2cm, align = center] at (3,1.5) {\small\textsf{OBS \\ Treated}};
\node[text width=2cm, align = center] at (6,1.5) {\small\textsf{OBS \\ Treated}};
\node[text width=2cm, align = center] at (9,1.5) {\small\textsf{OBS \\ Control}};

\node[text width=2cm, align = center] at (7.5,-4.5) {\Large\textsf{Right}};
\node[text width=2cm, align = center] at (1.5,-4.5) {\Large\textsf{Left}};

\end{tikzpicture}
\caption{A feasible flow $f(\cdot): \mathcal{E} \mapsto \{0, 1\}$ in a dense network. Thick, black lines correspond to edges with $f(e) = 1$ while light gray lines correspond to edges with $f(e) = 0$. The feasible flow yields a matched sample of three pairs: $\mathcal{M} = \{(\tau_1, \gamma_1), (\tau_2, \gamma_5), (\tau_3, \gamma_4)\}$.}
\label{fig: tripartite new scheme with feasible flow}
\end{figure}
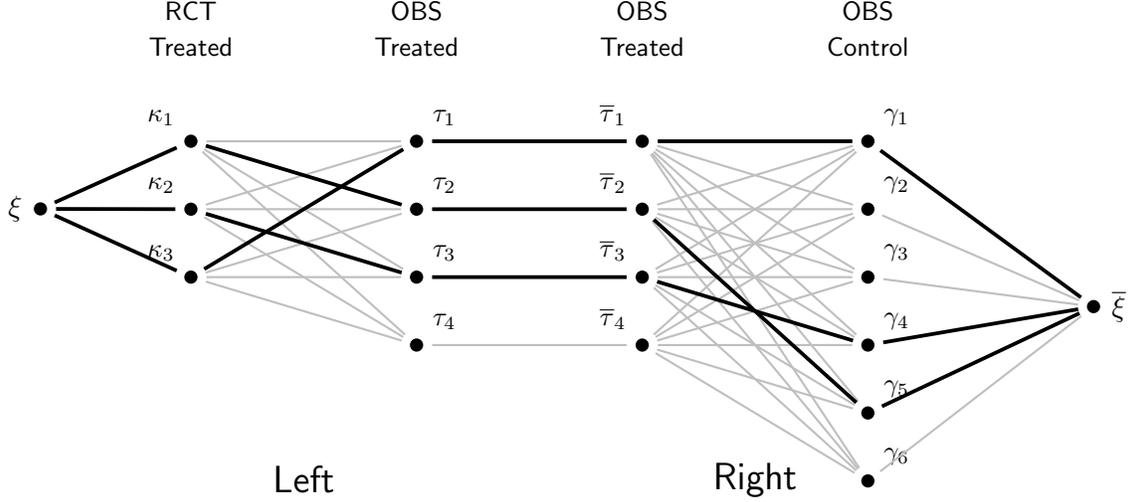

It is beneficial to consider a concrete example. The toy example in Figure \ref{fig: tripartite new scheme} has $R = 3$, $T = 4$, and $C = 6$. Consider setting $k = 1$ so that all edges in the network have capacity $1$. Thick, black lines in Figure \ref{fig: tripartite new scheme with feasible flow} correspond to one (out of ${4 \choose 3} \times 6 \times 5 \times 4 = 480$) feasible flows in this network. The left part of the network helps select OBS treated units using RCT units as a template, while the right part of the network performs the actual statistical matching and outputs matched samples. Formally, the matched samples $\mathcal{M}$ is defined by 
\begin{equation}
    \mathcal{M} = \left\{(\tau_t, \gamma_c) ~\text{such that}~f\{(\tau_t, \overline\tau_t)\} = f\{(\overline\tau_t, \gamma_c)\} = 1\right\}.
\end{equation}
For instance, the matched samples returned by the feasible flow in Figure \ref{fig: tripartite new scheme with feasible flow} consist of $\mathcal{M} = \{(\tau_1, \gamma_1), (\tau_2, \gamma_5), (\tau_3, \gamma_4)\}$.

\subsection{Basic Network Structure: Cost, Probability of Participation, Propensity Score}
\label{subsec: basic structure: cost}
While network infrastructure, i.e., vertices, edges, and capacities, determines the collection of all feasible flows, costs associated with each edge help select one best suited for empirical researchers' specific purposes. Let $\text{cost}(e)$ denote the cost associated with edge $e \in \mathcal{E}$. In the basic network structure depicted in Figure \ref{fig: tripartite new scheme}, we let $\text{cost}\{(\xi, \kappa_r)\} = \text{cost}\{(\tau_t, \overline\tau_t)\} = \text{cost}\{\gamma_c, \overline\xi\} = 0$ for $r = 1, \dots, R$, $t = 1, \dots, T$, and $c = 1, \dots, C$. 

Costs associated with edges $(\kappa_r, \tau_t)$ play an important role in forcing selected OBS samples to mimic RCT units in some covariates. Suppose that each RCT unit $\kappa_r$ is associated with a vector of covariates $\widetilde{\boldsymbol{x}}$ and each OBS unit $(\widetilde{\boldsymbol{x}}, \boldsymbol{x})$. As in the RCT DUPLICATE example, $\widetilde{\boldsymbol{x}}$ contains roughly $20$ covariates that both RCT and observational database collected while $\boldsymbol{x}$ consist of more than $100$ additional covariates available only in the claims database. Let $\delta_{\kappa_r, \tau_t} (\widetilde{\boldsymbol{x}})$ denote a measure of covariate distance between $\kappa_r$ and $\tau_t$ in $\widetilde{\boldsymbol{x}}$. The cost associated with an edge of the form $(\kappa_r, \tau_t)$ is equal to $\delta_{\kappa_r, \tau_t} (\widetilde{\boldsymbol{x}})$, i.e.,
\begin{equation}
    \text{cost}\{(\kappa_r, \tau_t)\} = \delta_{\kappa_r, \tau_t} (\widetilde{\boldsymbol{x}}).
\end{equation}
 One intriguing strategy equals $\delta_{\kappa_r, \tau_t} (\widetilde{\boldsymbol{x}})$ to a scalar ``balancing score" of $\widetilde{\boldsymbol{x}}$ (\citealp{rosenbaum1983central}), so that matching on this balancing score stochastically balances $\widetilde{\boldsymbol{x}}$. In the context of generalizing RCT's effect estimates to a target population, \citet{stuart2011use} defined and studied the conditional probability of selecting into the RCT group rather than the OBS group, which is referred to as the ``probability of participation." In practice, researchers could collect the RCT treated units $\mathcal{K}$ and OBS treated units $\mathcal{T}$, and estimate the conditional probability of being selected into the RCT given covariates $\widetilde{\boldsymbol{x}}$ using a logistic regression (\citealp{stuart2011use}).

Lastly, let $\Delta_{\tau_t, \gamma_c}\{(\widetilde{\boldsymbol{x}}, \boldsymbol{x})\}$ denote a measure of distance between OBS treated unit $\tau_t$ and OBS control unit $\gamma_c$ in their observed covariates $(\widetilde{\boldsymbol{x}}, \boldsymbol{x})$. The cost associated with an edge $(\overline\tau_t, \gamma_c)$ is set to $\Delta_{\tau_t, \gamma_c}\{(\widetilde{\boldsymbol{x}}, \boldsymbol{x})\}$, i.e.,
\begin{equation}
    \text{cost}\{(\overline\tau_t, \gamma_c)\} = \Delta_{\tau_t, \gamma_c}\{(\widetilde{\boldsymbol{x}}, \boldsymbol{x})\}.
\end{equation}
Different specifications of $\Delta_{\tau_t, \gamma_c}\{(\widetilde{\boldsymbol{x}}, \boldsymbol{x})\}$ have been extensively studied in the literature (\citealp{rosenbaum2002observational,rosenbaum2010design, stuart2010matching}). Perhaps the most widely used strategy is to equal $\Delta_{\tau_t, \gamma_c}\{(\widetilde{\boldsymbol{x}}, \boldsymbol{x})\}$ to the absolute difference in \citet{rosenbaum1983central}'s propensity score. Alternatively, one may let $\Delta_{\tau_t, \gamma_c}\{(\widetilde{\boldsymbol{x}}, \boldsymbol{x})\}$ represent the Mahalanobis distance in $(\widetilde{\boldsymbol{x}}, \boldsymbol{x})$ (\citealp{cochran1973controlling,rubin1979using}), or the Mahalanobis distance within a propensity score caliper (\citealp{rosenbaum1985constructing}). 

\subsection{Minimum Cost Flow, Complexity, Trade-Off Between Internal and External Validity}
\label{subsec: tradeoff}
The cost of a feasible flow $f(\cdot)$ in the proposed network is equal to
\begin{equation}
\label{eq: cost of a feasible flow}
    \text{cost}(f) = \sum_{e \in \mathcal{E}} f(e)\cdot \text{cost}(e) = \underbrace{\sum_{\substack{(\kappa_r, \tau_t) \in \mathcal{E}: \\f\{(\kappa_r, \tau_t)\} = 1}} \delta_{\kappa_r, \tau_t} (\widetilde{\boldsymbol{x}})}_{S1} ~+ \underbrace{\sum_{\substack{(\overline\tau_t, \gamma_c) \in \mathcal{E}: \\f\{(\overline\tau_t, \gamma_c)\} = 1}} \Delta_{\tau_t, \gamma_c}\{(\widetilde{\boldsymbol{x}}, \boldsymbol{x})\}}_{S2},
\end{equation}
and a flow $f(\cdot)$ is a minimum cost flow if it is feasible and every other feasible flow has a cost at least as high as $\text{cost}(f)$. A minimum cost flow can be found in $O(|\mathcal{V}|\cdot|\mathcal{E}| + |\mathcal{V}|^2\log(|\mathcal{V}|))$ operations (\citealp{korte2011combinatorial}) and in the proposed network, we have $|\mathcal{V}| = O(C)$, $|\mathcal{E}| = O(C^2)$, so that the computation complexity $O(|\mathcal{V}|\cdot|\mathcal{E}| + |\mathcal{V}|^2\log(|\mathcal{V}|))$ simplifies to $O(C^3)$. We discuss ways to sparsify the network and speed up computation in Supplemental Material B. In the statistical computing software $\textsf{R}$, the minimum cost flow can be found via \citet{bertsekas1991linear}'s auction algorithm implemented by \citet{bertsekas1988relax} and made available by \citet{hansen2007optmatch} and \citet{pimentel2015large}.

There is a tension between internal and external validity in a matched observational study. Take the RCT DUPLICATE initiative as an example. In their emulation of the SAVOR-TIMI 53 trial using observational data, \citet{franklin2021emulating} matched closely on and balanced almost $120$ observed covariates; however, the effect estimate obtained from observational data (HR: $0.81$; $95\%$ CI: $(0.76, 0.86)$) was significantly different from the RCT effect estimate (HR: $1.00$; $95\%$ CI: $(0.89, 1.12)$). There are many possible explanations in this large discrepancy, one of which is a quite large difference in baseline characteristics of the RCT versus OBS populations, including a disparity in some important preexisting conditions like MI. It is one possibility that both the observational study and RCT effect estimates are internally valid; they are just reporting effects for different populations.

According to our formulation, the cost $S1 = \sum_{\substack{(\overline\tau_t, \gamma_c) \in \mathcal{E}: \\f\{(\overline\tau_t, \gamma_c)\} = 1}} \Delta_{\tau_t, \gamma_c}\{(\widetilde{\boldsymbol{x}}, \boldsymbol{x})\}$ in expression \eqref{eq: cost of a feasible flow} measures the homogeneity between the matched treated and control groups. Well-matched samples have a small $S1$ value and better internal validity. On the other hand, the cost $S2 = \sum_{\substack{(\kappa_r, \tau_t) \in \mathcal{E}: \\f\{(\kappa_r, \tau_t)\} = 1}} \delta_{\kappa_r, \tau_t} (\widetilde{\boldsymbol{x}})$ measures how much the matched samples mimic a target population, and a small $S2$ corresponds to improved generalizability to the target population. 

To facilitate exploring the trade-off between a matched cohort study's internal validity and its generalizability to a target population, we replace $\Delta_{\tau_t, \gamma_c}\{(\widetilde{\boldsymbol{x}}, \boldsymbol{x})\}$ with $\lambda\cdot\Delta_{\tau_t, \gamma_c}\{(\widetilde{\boldsymbol{x}}, \boldsymbol{x})\}$ for some $\lambda > 0$ so that $\text{cost}(f)$ becomes:
\begin{equation}
\label{eq: cost of a feasible flow weighted}
   \text{cost}(f) = \sum_{e \in \mathcal{E}} f(e)\cdot \text{cost}(e) = \sum_{\substack{(\kappa_r, \tau_t) \in \mathcal{E}: \\f\{(\kappa_r, \tau_t)\} = 1}} \delta_{\kappa_r, \tau_t} (\widetilde{\boldsymbol{x}}) ~+ \lambda\cdot\bigg\{\sum_{\substack{(\overline\tau_t, \gamma_c) \in \mathcal{E}: \\f\{(\overline\tau_t, \gamma_c)\} = 1}} \Delta_{\tau_t, \gamma_c}\{(\widetilde{\boldsymbol{x}}, \boldsymbol{x})\}\bigg\}.
\end{equation}
According to this formulation, a large $\lambda$ value gives priority to a matched cohort study's internal validity, while a small $\lambda$ value prioritizes its generalizability to the target population. A similar weighting scheme is also used in \citet{zhang2021matching} but for a different purpose.

In Web Appendix B, we further discuss how to speed up computation and incorporate additional design features, including exact and near-exact matching on effect modifiers, \citeauthor{rosenbaum1989optimal}'s fine and near-fine balance \citeyearpar{rosenbaum1989optimal} and how to force including or excluding certain treated units, by modifying various aspects of the proposed network structure.

\subsection{Practical considerations and software availability}
\label{subsec: practical considerations}
The proposed matching algorithm is integrated in the package \textsf{match2C} available via the statistical computing software \textsf{R} (\citealp{R_software}) with a detailed tutorial.

To successfully implement the method, users need to specify two parameters. Parameter $k$ controls the size of the matched cohort and is determined by the size of the template, size of the observational database, overlap of the template and observational database in covariate distributions and power considerations. In general, a small $k$ value corresponds to constructing a small matched cohort and this small matched cohort by design will be more closely matched and better resemble the template compared to a larger matched cohort corresponding to a larger $k$ value; see the small illustrative example in Web Appendix C.2 and simulation studies in Section \ref{sec: simulation and software}. If researchers find the matched comparison not adequately powered, then larger $k$ values should be explored. Parameter $\lambda$ controls the trade-off between the validity of the matched comparison and its generalizability to the target population. By default, the parameter $\lambda$ is set to a large number to prioritize the validity of the matched cohort study. When the treated and control groups in the observational database are well-overlapped so that there are many possible internally-valid matched cohort studies, researchers may then reduce $\lambda$ to further improve the matched cohort study's generalizability.

Matching is part of the design of an observational study and should be carried out without looking at the outcome data. Good practice includes keeping time-stamped analysis logs for review and posting a detailed pre-analysis protocol; see, e.g., \cite{franklin2020nonrandomized, franklin2021emulating}. Provided that no outcome data are viewed, researchers typically perform statistical matching multiple times and select the design based on covariate balance. Recently, many formal diagnostic tests of covariate balance have been proposed \citep{gagnon2019classification}. Researchers could perform a formal diagnostic test, e.g., \citeauthor{gagnon2019classification}'s \citeyearpar{gagnon2019classification} classification permutation test (CPT) to see if there is any residual imbalance in observed covariates $(\widetilde{\boldsymbol{x}}, \boldsymbol{x})$ between two groups in the matched cohort; analogously, a formal test could be carried out to examine residual imbalance in common covariates $\widetilde{\boldsymbol{x}}$ between the matched group and the template. We will discuss these aspects more concretely when examining the WHI study in Section \ref{sec: WHI real data}.

\section{Simulation study}
\label{sec: simulation and software}

\subsection{Goal; structure; measurement of success}
Our primary goal in this section is to examine how the study design delivered by different matching algorithms affects the bias. In particular, we are interested in the case where the treatment effect is heterogeneous and the distributions of effect modifiers vary in the observational database and the target population. We considered the following target population: $\mathbf{X} \sim \text{Multivariate Normal}\left(\mathbf\mu, \mathbf \Sigma\right)$ with $\mathbf \mu = (0.25, 0, 0, 0, 0)^{\text{T}}$ and $\mathbf\Sigma = \mathbf I_{5\times 5}$, and generated a template $\mathcal{K}$ consisting of a random sample of size $300$ from the target population. We consider an observational database with $|\mathcal{T}| = 1,000$ treated and $|\mathcal{C}| = 3,000$ control units. The data-generating process for units in the observational database and statistical matching procedures to be investigated are specified via the following factorial design:
\begin{description}
\item \textbf{Factor 1:} Dimension of covariates in the observational database, $d$: $10$, $30$, and $50$.
\item \textbf{Factor 2:} Overlap, $\theta$: 
$\mathbf{X} \sim \text{Multivariate Normal}\left(\mathbf\mu, \mathbf \Sigma\right)$, with $\mathbf \mu = (\theta Z, 0, \dots, 0)^{\text{T}}$ and $\mathbf\Sigma = \mathbf I_{d\times d}$. We consider $\theta = 0.50$ and $0.75$.

\item \textbf{Factor 3:} Matching algorithms to be investigated, $\mathcal{M}$: 
\begin{enumerate}
    \item  $\mathcal{M}_{\textsf{opt}}$: matching according to two criteria \citep{zhang2021matching}: (i) minimizing the earthmover's distance, a measure of distance between two probability distributions \citep{levina2001earth}, between the distributions of the estimated propensity scores in the treated and matched control groups, and (ii) minimizing the within-matched-pair robust Mahalanobis distances. Algorithm $\mathcal{M}_{\textsf{opt}}$ produces $1,000$ matched pairs and does not make use of the template.
    
    \item $\mathcal{M}_{\textsf{template},~k = 1}$: matching according to the proposed network structure (Figure \ref{fig: tripartite new scheme with feasible flow}) with $k = 1$. Algorithm $\mathcal{M}_{\textsf{template},~k = 1}$ produces $300$ matched pairs.
   
    \item $\mathcal{M}_{\textsf{template},~k = 2}$: similar to $\mathcal{M}_{\textsf{template},~k = 1}$ but with $k = 2$. Algorithm $\mathcal{M}_{\textsf{template},~k = 2}$ produces $600$ matched pairs.
\end{enumerate}

\item \textbf{Factor 4:} Tuning parameter in $\mathcal{M}_{\textsf{template},~k}$, $\lambda$: $100$, $1$, $0.01$.
\end{description}

Factor $1$ and $2$ define the data-generating process for units in the observational database. Factor $3$ and $4$ define a total of $1 + 2 \times 3 = 7$ matching algorithms to be investigated. As discussed in Section \ref{subsec: tradeoff}, the tuning parameter $\lambda$ controls the trade-off between a matched cohort study's validity and its generalizability to the target population.

For each unit, we further generate two potential outcomes:
\begin{equation}
    \label{eq: simu outcome}
    Y(0) \sim \text{Normal}(\mathbf{X}^T\boldsymbol{\nu}, 1),\qquad Y(1) = Y(0) + \beta(X_1),
\end{equation}
and the observed outcome satisfies $Y = Z\cdot Y(1) + (1-Z)\cdot Y(0)$. Factor $5$ specifies the mean of the potential outcome $Y(0)$ and Factor $6$ specifies the treatment effect:
\begin{description}
\item \textbf{Factor 5:} Mean of $Y(0)$, $\mathbf{X}^T\boldsymbol{\nu}$: a constant $\boldsymbol{\nu}$ vector with all entries equal to $0$, $0.05$ or $0.1$.
\item \textbf{Factor 6:} Treatment effect, $\beta(X_1)$: a constant effect $\beta(X_1) = 2$, a mildly heterogeneous effect $\beta(X_1) = 2 - 0.2X_1$, and a strongly heterogeneous effect $\beta(X_1) = 2 - X_1$. 
\end{description}
The average treatment effect of the target population satisfies $\textsf{ATE}_{\text{target}} = 2$ when $\beta(X_1) = 2$, $\textsf{ATE}_{\text{target}} = 1.95$ when $\beta(X_1) = 2 - 0.2X_1$ and $\textsf{ATE}_{\text{target}} = 1.75$ when $\beta(X_1) = 2 - X_1$.

There are multiple ways to analyze matched-pair data. Examples include a parametric t-test, randomization inference \citep{rosenbaum2002observational,rosenbaum2010design} and regression adjustment \citep{rubin1979using, ho2007matching}. In this simulation study, we calculated a difference-in-means estimator for matched data produced by each of the $7$ algorithms and compared these $7$ effect estimates to the target parameter $\textsf{ATE}_{\text{target}}$ in data-generating process.

\subsection{Simulation results}
Table \ref{tbl: simulation results b = 0.5} summarizes the percentage of bias of each difference-in-means estimator when the overlap parameter $\theta = 0.50$. Web Appendix D.1 reports similar results for $\theta = 0.75$.


We have observed a few consistent trends. First, when the treatment effect is constant, the bias is relatively small under all data-generating processes and statistical matching algorithms under consideration. Second, when the treatment effect is heterogeneous and the effect modifier $X_1$ has a different distribution in the template and the group of treated units, the effect estimate obtained from $1000$ matched pairs constructed using algorithm $\mathcal{M}_{\textsf{opt}}$ is clearly biased from $\textsf{ATE}_{\text{target}}$, and the percentage of bias increases (i) as the distributions of $X_1$ in the target population and in the group of OBS treated units become increasingly dissimilar, that is, as $\theta$ increases, and (ii) as effect modification becomes more pronounced, that is, from $\beta(X_1) = 2$ (constant) to $\beta(X_1) = 2 - 0.2X_1$ (mild) to $\beta(X_1) = 2 - X_1$ (strong). In the most adversarial scenario with $\beta(X_1) = 2 - X_1$, the percentage bias of $\widehat{\theta}_{\mathcal{M}_{\textsf{opt}}}$ can be as large as $25\%$. We need to stress that although $\widehat{\theta}_{\mathcal{M}_{\textsf{opt}}}$ may not be generalized to the target population, it is an internally-valid estimator for the average treatment effect on the treated.

Our proposed algorithm outperforms $\mathcal{M}_{\textsf{opt}}$ in bias reduction against $\textsf{ATE}_{\text{target}}$ in all $6$ different implementations under all data-generating processes considered in this simulation study, although the gain in bias reduction differs from implementation to implementation. In particular, we observe that the gain is most pronounced when (i) $k$ is small so that a smaller treated group bearing more resemblance to the target template is constructed, and (ii) $\lambda$ is small so that the matching algorithm gives priority to resemblance to the template. 

\begin{table}[ht]
\centering
\caption{Percentage of bias with respect to $\textsf{ATE}_{\text{target}}$ of $7$ difference-in-means estimator constructed from matched samples obtained from each of the $7$ matching algorithms under consideration. The overlap parameter $\theta$ in \textbf{Factor 2} is equal to $0.50$.  Each cell is averaged over $1000$ simulations.}
\label{tbl: simulation results b = 0.5}
\resizebox{\textwidth}{!}{
\begin{tabular}{cccccccccccc}\hline \\ [-0.8em]\multirow{3}{*}{\begin{tabular}{c}Heterogeneity \\ Level $\beta(X_1)$\end{tabular}}
&\multirow{3}{*}{\begin{tabular}{c}DGP of $Y(0)$\\$\nu$ \end{tabular}}&& \multirow{3}{*}{\begin{tabular}{c}$\mathcal{M}_{\textsf{opt}}$\end{tabular}} 
&& \multicolumn{6}{c}{$\mathcal{M}_{\textsf{template}}$} \\ \\ [-0.8em] \cline{6-11} \\ [-0.8em]
&&&& & \multirow{2}{*}{\begin{tabular}{c}$k = 1$ \\ $\lambda = 0.01$ \end{tabular}}  
& \multirow{2}{*}{\begin{tabular}{c}$k = 1$ \\ $\lambda = 1$ \end{tabular}} 
& \multirow{2}{*}{\begin{tabular}{c}$k = 1$ \\ $\lambda = 100$ \end{tabular}} 
& \multirow{2}{*}{\begin{tabular}{c}$k = 2$ \\ $\lambda = 0.01$ \end{tabular}}  
& \multirow{2}{*}{\begin{tabular}{c}$k = 2$ \\ $\lambda = 1$ \end{tabular}} 
& \multirow{2}{*}{\begin{tabular}{c}$k = 2$ \\ $\lambda = 100$ \end{tabular}} \\ \\ \\ \\ [-0.8em]
    \multicolumn{11}{c}{$d = 10$} \\ \\ [-0.8em]
    
    \multirow{3}{*}{Constant} 
  &0    && 0.07\% && 0.21\% & -0.06\% & -0.06\% & -0.01\% & 0.05\% & -0.06\% \\ 
  &0.05 && 0.00\% && 0.33\% & 0.18\% & 0.14\% & 0.23\% & 0.25\% & 0.44\% \\ 
  &0.10 && 0.06\% && 0.59\% & 0.30\% & 0.44\% & 0.73\% & 0.66\% & 0.61\% \\  \\ [-0.8em]

       \multirow{3}{*}{Mild} 
  &0 && -5.17\% && -1.23\% & -1.19\% & -2.04\% & -1.52\% & -1.56\% & -2.56\% \\ 
  &0.05 && -4.80\% && -0.11\% & -0.15\% & -1.25\% & -0.53\% & -0.65\% & -1.77\% \\ 
  &0.10 && -5.00\% && 0.04\% & -0.20\% & -1.22\% & -0.58\% & -0.48\% & -1.44\% \\  \\ [-0.8em]

  \multirow{3}{*}{Strong} 
  &0 && -24.99\% && -4.01\% & -3.75\% & -9.05\% & -6.45\% & -5.96\% & -11.68\% \\ 
  & 0.05 && -25.18\% && -3.10\% & -2.92\% & -8.40\% & -5.50\% & -5.60\% & -11.43\% \\ 
  & 0.10 && -25.23\% && -3.68\% & -3.39\% & -8.07\% & -6.14\% & -5.91\% & -11.25\% \\ \\ [-0.8em]
     \multicolumn{11}{c}{$d = 30$} \\  \\ [-0.8em]
          \multirow{3}{*}{Constant} 
  &0 && 0.23\% && 0.09\% & 0.45\% & 0.40\% & 0.31\% & 0.33\% & 0.28\% \\ 
  &0.05 && 0.13\% && 1.04\% & 1.27\% & 1.37\% & 1.13\% & 1.19\% & 1.29\% \\ 
  &0.10 && 0.12\% && 1.55\% & 1.43\% & 1.62\% & 1.70\% & 1.61\% & 1.83\% \\  \\ [-0.8em]
     \multirow{3}{*}{Mild} 
     &0 && -4.77\% && -0.57\% & -1.32\% & -2.06\% & -1.07\% & -1.76\% & -2.50\% \\ 
     & 0.05 && -5.07\% && -0.55\% & -0.83\% & -1.56\% & -0.67\% & -1.27\% & -1.88\% \\ 
    & 0.10 && -4.63\% && 0.99\% & 0.80\% & 0.48\% & 0.62\% & 0.06\% & -0.41\% \\  \\ [-0.8em]
   \multirow{3}{*}{Strong} 
   &0 && -24.71\% && -3.57\% & -5.90\% & -9.28\% & -5.65\% & -9.50\% & -12.83\% \\ 
  &0.05 && -24.89\% && -3.13\% & -5.44\% & -8.80\% & -5.30\% & -8.96\% & -12.37\% \\ 
  &0.10 && -24.78\% && -2.49\% & -4.88\% & -8.01\% & -4.51\% & -8.21\% & -11.35\% \\  \\ [-0.8em]
  \multicolumn{11}{c}{$d = 50$} \\  \\ [-0.8em]
   \multirow{3}{*}{Constant} 
   &0 && 0.06\% && 0.31\% & -0.21\% & -0.21\% & 0.19\% & -0.12\% & -0.06\% \\ 
  &0.05 && 0.07\% && 0.77\% & 1.15\% & 1.12\% & 0.99\% & 1.29\% & 1.29\% \\ 
  &0.10 && -0.05\% && 1.33\% & 1.82\% & 2.07\% & 1.66\% & 2.02\% & 2.18\% \\ \\ [-0.8em]
   \multirow{3}{*}{Mild} 
   &0 && -5.12\% && -1.02\% & -1.59\% & -2.20\% & -1.60\% & -2.43\% & -2.94\% \\ 
  &0.05 && -5.15\% && 0.37\% & -0.69\% & -1.26\% & -0.37\% & -1.16\% & -1.63\% \\ 
  &0.10 && -5.03\% && 0.78\% & 0.20\% & -0.24\% & 0.47\% & 0.01\% & -0.40\% \\  \\ [-0.8em]
   \multirow{3}{*}{Strong} 
   &0 && -24.64\% && -3.48\% & -6.82\% & -9.48\% & -5.84\% & -10.26\% & -12.72\% \\ 
  &0.05 && -25.00\% && -3.22\% & -6.76\% & -9.06\% & -5.39\% & -9.88\% & -12.21\% \\ 
  &0.10 && -24.78\% && -1.80\% & -5.01\% & -7.37\% & -3.97\% & -8.34\% & -10.63\% \\ \\ [-0.8em] \hline
\end{tabular}}
\end{table}

\subsection{Additional simulation studies}
In Web Appendix D.2, we designed a simulation study to investigate the role of unmeasured confounding. In the presence of unmeasured confounding in the observational database, the bias consisted of two parts, a generalization bias and an unmeasured confounding bias. By removing most of the generalization bias using our proposed algorithm, researchers could better focus on the unmeasured confounding bias. In Web Appendix D.3, we further compared the computation cost of a network-flow-based algorithm and an MIP-based algorithm implemented in the \textsf{R} package \textsf{designmatch} based on an open source optimization routine GLPK. We found that the network-flow-based method largely outperforms the MIP-based method in this comparison, although we note that the performance of MIP-based methods would improve when they are implemented using a more powerful commercial optimization routine like \textsf{Gurobi}.

\section{Revisiting the Women’s Health Initiative (WHI) study}
\label{sec: WHI real data}
\subsection{Background and our goal}
The Women's Health Initiative (WHI) is a combined clinical trial and observational study. Postmenopausal women were screened for clinical trial eligibility; those who were ineligible or unwilling to participate in the trial were enrolled in the observational study. The design of the WHI study is described in \citet{study1998design}. One important goal of the WHI clinical trial is to evaluate the hypothesized cardioprotective effect of postmenopausal hormone therapy, following a substantial body of evidence from observational studies \citep{study1998design}. The WHI estrogen-plus-progestin (E + P for short) trial found a rather surprising elevation in coronary heart disease risk (\citealp{writing2002risks}) and sparked a lot of discussion regarding the discrepancy between clinical trial and observational study results. 

Many authors have speculated on why results would differ dramatically between the trial and observational study. Some major concerns include: (i) potential bias due to unmeasured confounding in observational studies \citep{humphrey2002postmenopausal}; (ii) biological differences between trial participants and those in the observational study \citep{Michels2003editorial} and (iii) differences in time since menopause at hormone therapy initiation \citep{prentice2005combined,willett2006re, hernan2008observational, prentice2009benefits}, among others. Table \ref{tb: WHI baseline characteristics} summarizes some important baseline covariates in the WHI trial and associated observational study and illustrates some of these concerns. Compared to past and never users, current users of estrogen-plus-progestin in the WHI observational study are younger, less likely to be black or Hispanic, more educated, and have more physical activity episodes per week. On the other hand, baseline covariates of participants in the control and intervention groups are similar in the WHI trial by virtue of randomization. There is also a tangible difference in socioeconomic status and smoking status between trial and observational study participants. Moreover, trial participants initiated their HRT at a much older age compared to the current users in the observational study.

\begin{table}[ht]
\caption{Important baseline characteristics of the WHI observational study and WHI trial subjects. Mean (SE) are reported for continuous variables and count (\%) for categorical variables.}
\label{tb: WHI baseline characteristics}
\centering
\resizebox{\textwidth}{!}{
\begin{tabular}{lcccccc}
  \hline \\ [-0.8em]
  &\multicolumn{2}{c}{WHI Observational Data} &&\multicolumn{2}{c}{WHI Trial}\\ \\ [-0.8em] \cline{2-3}\cline{5-6}
  &\multirow{4}{*}{\begin{tabular}{c}\textbf{Never/Past} \\ \textbf{Users} \\($\mathbf{n = 75303}$)\end{tabular}}
  & \multirow{4}{*}{\begin{tabular}{c}\textbf{Current} \\ \textbf{Users} \\($\mathbf{n = 18340}$)\end{tabular}}&
  & \multirow{4}{*}{\begin{tabular}{c} \textbf{Control} \\($\mathbf{n = 8102}$)\end{tabular}}
  & \multirow{4}{*}{\begin{tabular}{c}\textbf{E+P} \\ \textbf{Intervention} \\($\mathbf{n = 8506}$)\end{tabular}}\\ \\ \\ \\
  
  \textbf{Age at screening} &  64.30 (7.37) &  60.84 (6.69) &&  63.33 (7.11) &  63.23 (7.13) \\ 
  \textbf{Race/Ethnicity}  \\ 
  \hspace{0.5cm}White &  61704 (81.9)  &  16285 (88.8)  &&   6805 (84.0)  &   7141 (84.0) \\ 
  \hspace{0.5cm}Black/Hispanic &  10166 (13.5)  &   1074 ( 5.9)  &&    989 (12.2)  &   1019 (12.0) \\ 
  \hspace{0.5cm}Other &   3433 ( 4.6)  &    981 ( 5.3)  &&    308 ( 3.8)  &    346 ( 4.1)   \\ 
  \textbf{Education} \\ 
  \hspace{0.5cm}College or above &  30418 (40.4)  &  10044 (54.8)  &&   3011 (37.2)  &   3111 (36.6)  \\ 
  \hspace{0.5cm}Some college &  28074 (37.3)  &   5853 (31.9)  &&   3060 (37.8)  &   3357 (39.5) \\ 
  \hspace{0.5cm}High school diploma/GED &  13068 (17.4)  &   2050 (11.2)  &&   1609 (19.9)  &   1615 (19.0) \\ 
  \hspace{0.5cm}Other &    3743 ( 5.0)  &    393 (2.1)  &&     422 (5.2)  &  423 (5.0) \\ 

  \textbf{Blood Pressure} \\
  \hspace{0.5cm}Systolic & 128 (18.12) & 123 (16.92) && 128 (17.53) & 128 (17.63)  \\ 
  \hspace{0.5cm}Diastolic &  75 (9.42) &  74 (8.99) &&  76 (9.09) &  76 (9.12)   \\ 
  \textbf{BMI} &  27.62 (5.98) &  25.84 (5.15) &&  28.50 (5.91) &  28.46 (5.82)   \\ 
  \textbf{Smoking} \\ 
  \hspace{0.5cm}NA &   1004 ( 1.3)  &    220 ( 1.2)  &&     98 ( 1.2)  &     83 ( 1.0) \\ 
  \hspace{0.5cm}Current Smoker &   4850 ( 6.4)  &    940 ( 5.1)  &&    838 (10.4)  &    880 (10.3)   \\ 
  \hspace{0.5cm}Never Smoked &  38296 (50.9)  &   8710 (47.5)  &&   3999 (49.4)  &   4178 (49.1)   \\ 
  \hspace{0.5cm}Past Smoker &  31048 (41.3)  &   8448 (46.1)  &&   3157 (39.0)  &   3362 (39.5)   \\ 
  \textbf{No. of PA episodes} \\
  \hspace{0.5cm}Total &   5.28 (4.12) &   5.80 (4.12) &&   4.77 (4.06) &   4.74 (4.10)  \\ 
  \hspace{0.5cm}Medium to strenuous &   2.94 (3.37) &   3.59 (3.57) &&   2.58 (3.24) &   2.50 (3.21)   \\ 
   \textbf{HRT Initiation}  \\
  \hspace{0.5cm}NA &  67259 (89.3)  & 0 &&   6706 (82.8)  &      0   \\ 
  \hspace{0.5cm}Age at initiation   &  52.04 (7.35) &  53.87 (6.84) &&  53.57 (6.53) &  61.93 (8.15) \\ 
  \textbf{Previous E+P use in years} &   0.47 (2.12) &   7.02 (5.64) &&   0.64 (2.23) &   0.70 (2.34)   \\ \textbf{Unopposed estrogen use ever}            \\
  \hspace{0.5cm}Yes &32705 (43.4) & 2170 (11.8)  &&865 (10.7) &903 (10.6)\\
  \hspace{0.5cm}No &42598 (56.6)  &16170 (88.2)  &&7237 (89.3) &7603 (89.4)\\
  \textbf{Reproductive history}\\
  \hspace{0.5cm}No ovary removed &48303 (64.1) &16906 (92.2) & &7705 (95.1) &8083 (95.0)\\
   \hline
\end{tabular}}
\end{table}

\textcolor{black}{Our proposed matching algorithm could help alleviate these concerns. The first concern regarding the validity of the observational study could be alleviated if matched observational study participants are balanced for a large number of baseline covariates including detailed demographics, preexisting comorbid conditions and personal habits. The second and third concerns about the comparability between OBS and RCT participants in their cardiovascular risk profile and HRT initiation time may be mitigated by treating the RCT cohort as a template and constructing matched OBS pairs that resemble the RCT cohort in these aspects.}

\textcolor{black}{Our proposed algorithm cannot fully address the concern regarding unmeasured confounding in the observational study. Another concern that cannot be adequately addressed is the difference in estrogen-plus-progestin usage time prior to the WHI study between RCT and OBS samples:} current users in the observational study had used HRT for more than $7$ years on average, while participants of the trial had used HRT for less than a year on average. This difference is largely due to the design of the WHI study: the observational study enrolled a cross-section of post-menopausal women that included many who had been using the therapy for years (\citealp{willett2006re}) while most trial participants were never-users of the therapy by the time of randomization. This lack of overlap between RCT and OBS samples in previous estrogen-plus-progestin usage makes it virtually impossible to balance this aspect between RCT and OBS samples using any matching tool; nevertheless, \textcolor{black}{in our opinion, alleviating some major concerns helps researchers critically examine the remaining explanations and is a meaningful step towards reconciling the WHI RCT and OBS findings.}

\vspace{-0.1cm}
\subsection{Template and three matched samples}
Our goal is to construct well-matched pairs from the WHI observational study that resemble the WHI trial participants in the following sense: (i) matched pairs resemble the WHI trial participants in cardiovascular risk profile, and (ii) treated participants in the matched pairs resemble the WHI trial intervention group in their HRT initiation time. Our template consists of a random sample of $1,000$ WHI trial participants in the intervention group with the following covariates: risk factors listed in Table \ref{tb: WHI baseline characteristics} plus the HRT initiation time. 



We applied the proposed matching algorithm to constructing matched pairs of different sizes corresponding to choosing different $k$'s in Section \ref{subsec: basic structure: c and c}. Our desired matched OBS cohort would have similar cardiovascular risk profile as the template and are closely matched for many other baseline covariates to maximally guard against the unmeasured confounding bias due to the non-randomized nature of the WHI observational study. In addition to risk factors listed in Table \ref{tb: WHI baseline characteristics}, we further matched on participants' region, partners' education level, income level, marital status, reproductive history and eight important preexisting conditions.

Table \ref{tbl: matched samples I} summarizes results from two matched samples constructed using the proposed algorithm with different parameters. Match $\textsf{M1}$ constructed $10,000$ matched pairs of two observational study participants: one treated and the other control. Match $\textsf{M1}$ used a Mahalanobis distance with an estimated generalizability score caliper (caliper size = $0.05$) and a Mahalanobis distance with an estimated propensity score caliper (caliper size = $0.15$) on the left and right part of the network depicted in Figure \ref{fig: tripartite new scheme with feasible flow}, respectively. Because of the large sample size ($\approx 100,000$ in the observational study), we applied a ``hard" caliper in the sense that edges connecting one OBS treated and one OBS control units are removed whenever they differ in their estimated propensity score by more than the caliper size; in this way, the network is sparsified and computation is boosted; see Web Appendix B for details. We set $\lambda = 100$ to give priority to internal validity of the matched comparison. Match $\textsf{M2}$ is similar to $\textsf{M1}$ except that we only formed $3,000$ matched pairs. Lastly, match $\textsf{M0}$ formed $18,340$ matched pairs exhausting every treated observational study participant. The balance table and propensity score distributions of $\textsf{M0}$ can be found in Web Appendix E.

Judging from internal validity, all three matched cohorts are acceptable: the absolute standardized mean differences (SMDs) of most cardiovascular risk factors and additional OBS covariates are less than $0.1$, or one-tenth of one pooled standard deviation \citep{silber2001multivariate}, and many are below $0.01$. However, the three designs differ, sometimes significantly, in their resemblance to the trial population. In particular, design $\textsf{M0}$ differs from the trial participants in their cardiovascular risk profile (e.g., the percentage of black/Hispanic is $6\%$ in $\textsf{M0}$ compared to $12\%$ in the template) and HRT initiation age ($54$ in $\textsf{M0}$ compared to $62$ in the trial population). On the other hand, both $\textsf{M1}$ and $\textsf{M2}$ are similar to the trial population in the cardiovascular risk profile: the percentage of white (black/Hispanic) women was $85\%$ ($10\%$) in $\textsf{M1}$, $84\%$ ($12\%$) in $\textsf{M2}$ and $84\%$ ($12\%$) in the template, compared to $89\%$ ($6\%$) in $\textsf{M0}$; the percentage with a college degree or above is $41\%$ in $\textsf{M1}$, $36\%$ in $\textsf{M2}$, and $38\%$ in the template, compared to $55\%$ in $\textsf{M0}$. \textcolor{black}{In fact, the matched treated group in either $\textsf{M1}$ or $\textsf{M2}$ is now indistinguishable from the template based on the cardiovascular risk profile as judged by \citeauthor{gagnon2019classification}'s classification permutation test \citeyearpar{gagnon2019classification}.} Moreover, the treated group of $\textsf{M1}$ now initiated their HRT at an average age of $57$ with an average $5.54$-year of HRT usage prior to enrolling in the WHI study, and the treated group of $\textsf{M2}$ initiated their HRT at an average age of $61$ with an average of $3.24$-year previous HRT usage. The HRT initiation age in the design \textsf{M2} is now closer to that of the trial population, though a potentially meaningful discrepancy in the previous HRT usage persists ($3.24$ vs $0.70$ years).

\vspace{-0.15cm}
\subsection{Comparing survival outcomes}
\begin{figure}[ht]
   \centering
     \subfloat[Unadjusted]{\includegraphics[width = 0.49\columnwidth]{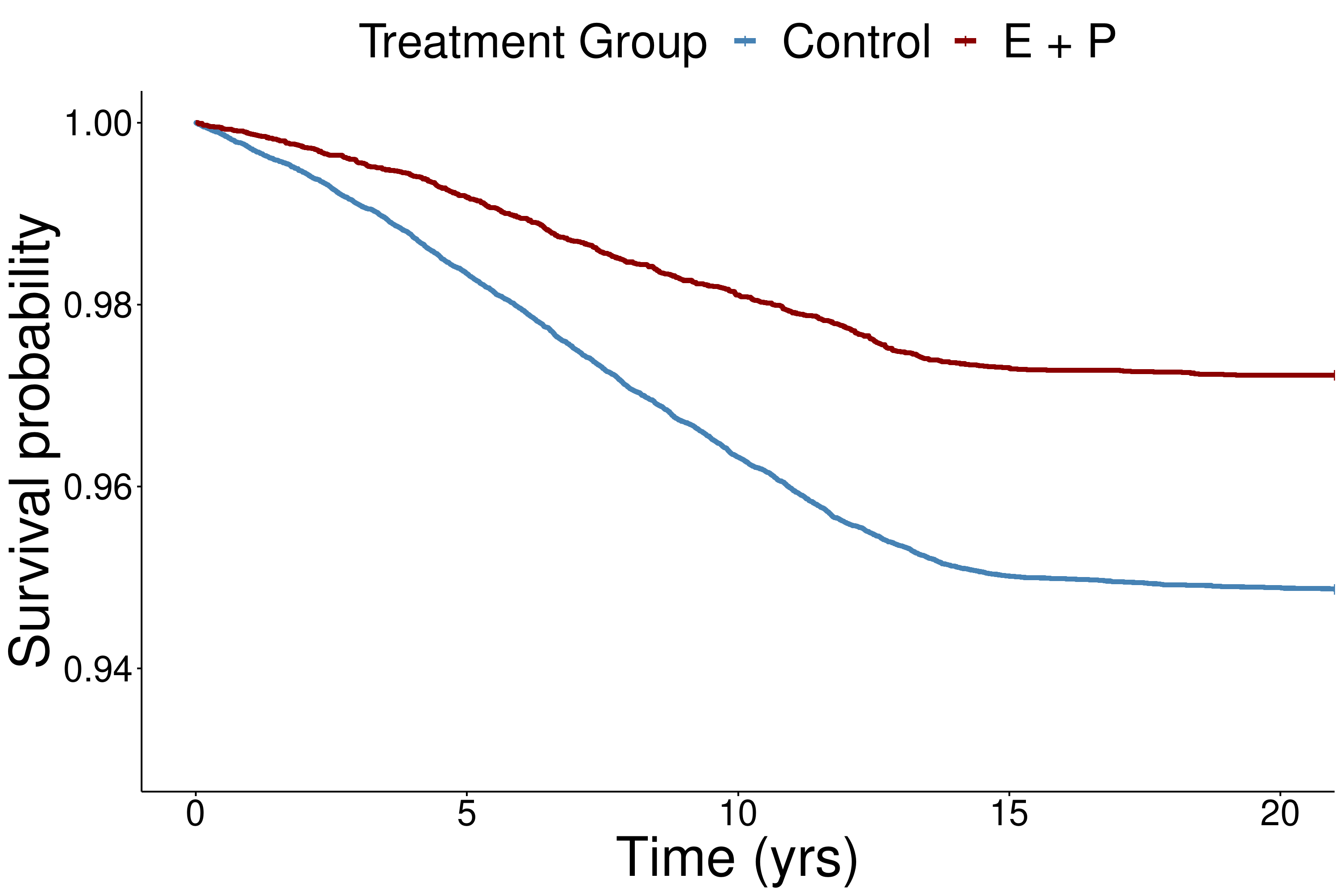}}
     \subfloat[Design $\textsf{M0}$]{\includegraphics[width = 0.49\columnwidth]{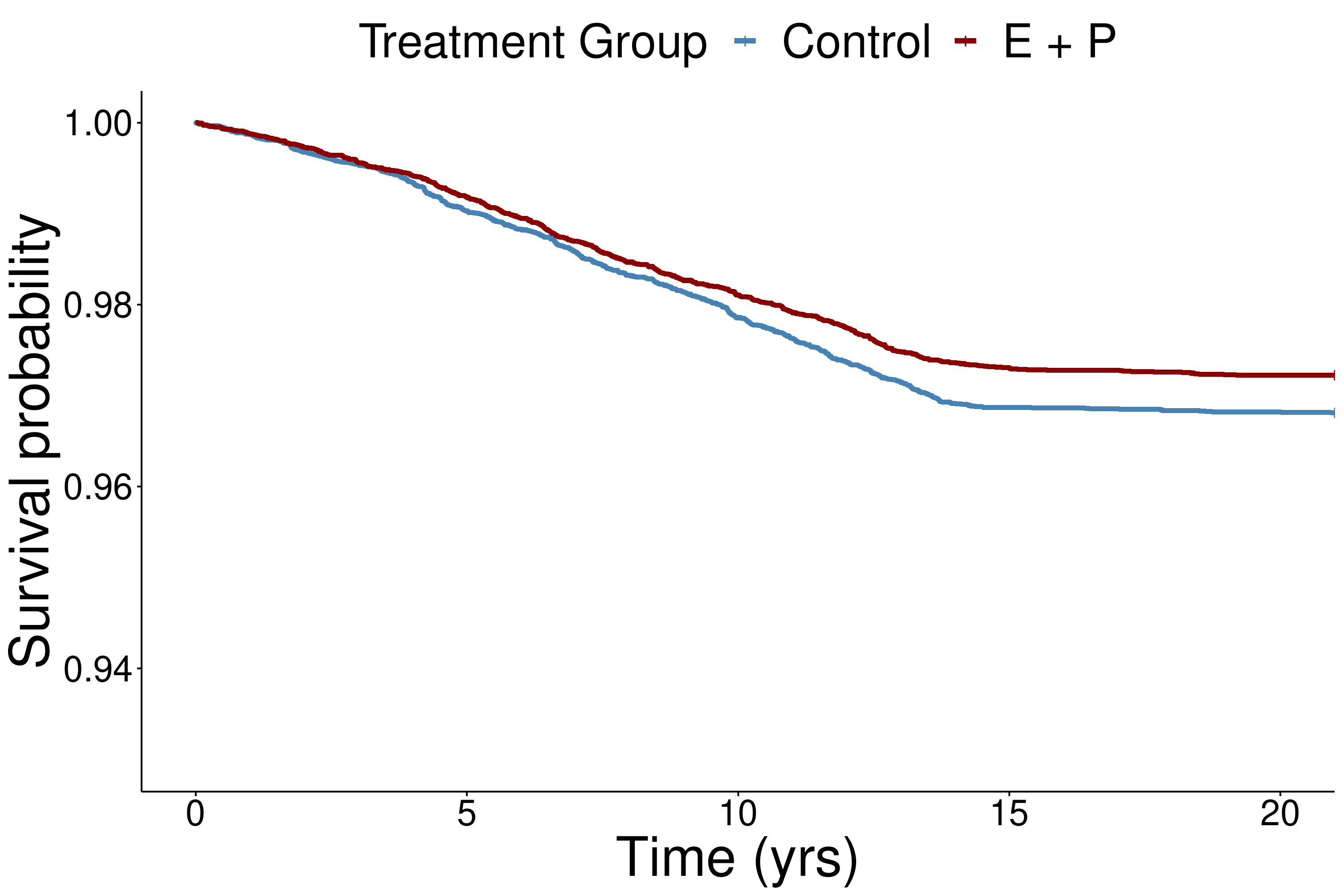}}\\
     \subfloat[Design $\textsf{M1}$]{\includegraphics[width = 0.49\columnwidth]{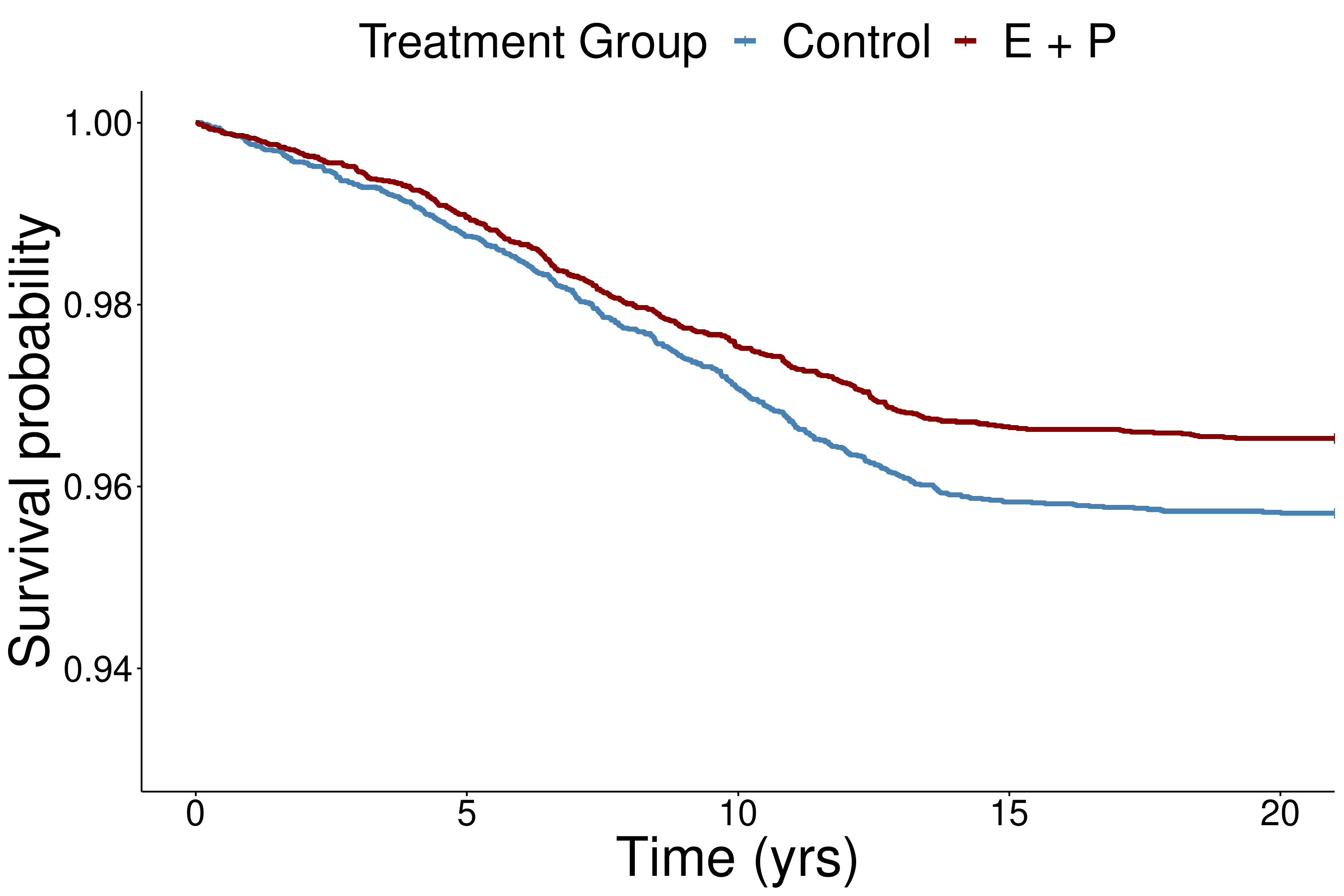}}
      \subfloat[Design $\textsf{M2}$]{\includegraphics[width = 0.49\columnwidth]{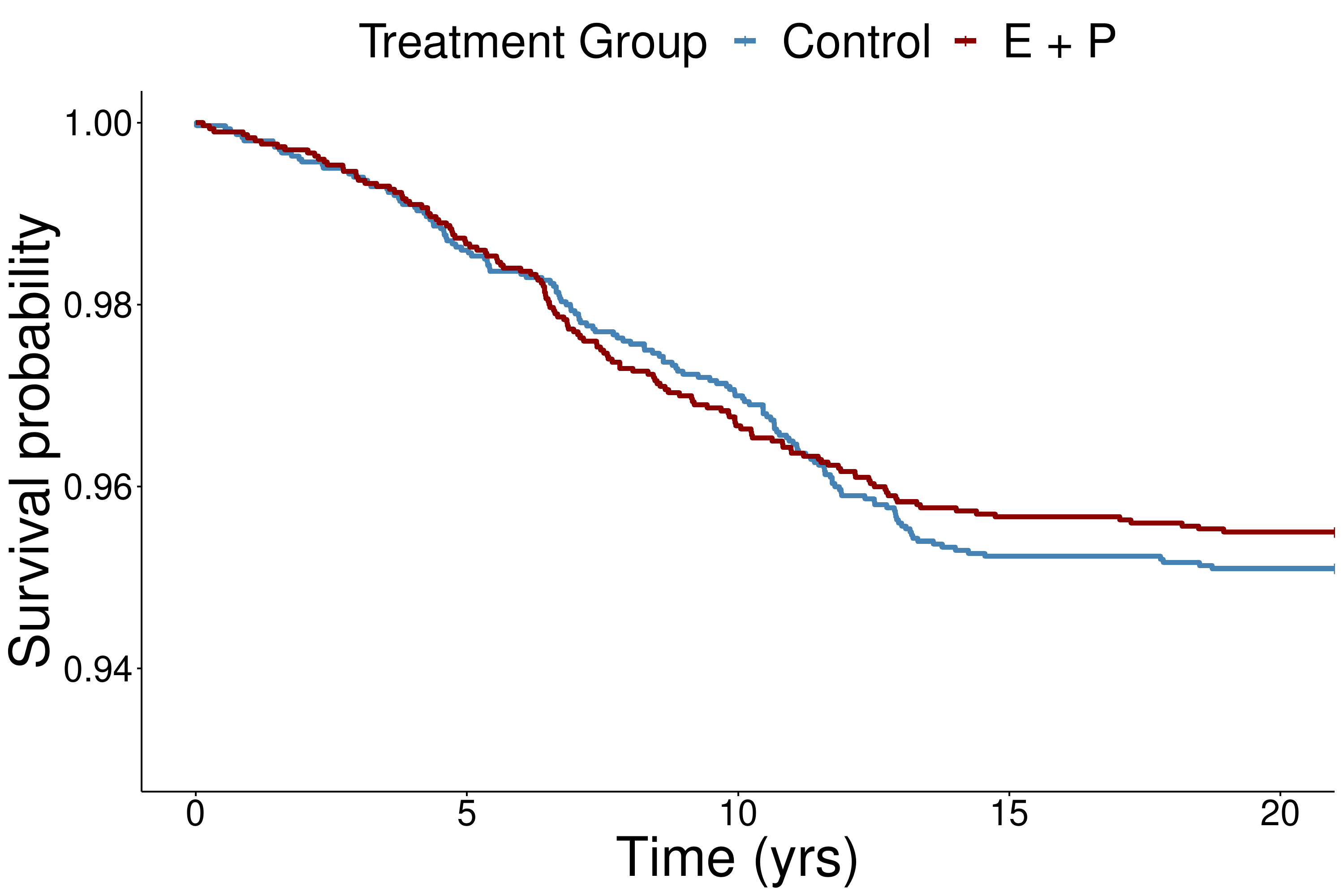}}
     \caption{Survival outcomes for coronary heart disease (CHD) in the unadjusted observational study samples (top left), design \textsf{M0} (top right), design \textsf{M1} (bottom left), and design \textsf{M2} (bottom right).}
\label{fig: real data K-M curve}
\end{figure}

Figure \ref{fig: real data K-M curve} plots the Kaplan-Meier curves (\citealp{kaplan1958nonparametric}) in the treatment and control groups in each study design. Compared to all controls in the unadjusted analysis (top left panel of Figure \ref{fig: real data K-M curve}), matched controls in \textsf{M0} have higher survival probability, reflecting the fact that these matched controls are now more identical to the treated OBS participants and healthier; however, the cardioprotective effect of HRT persists ($P$-value = $0.020$ according to \citeauthor{obrien1987paired}'s \citeyearpar{obrien1987paired} matched-pair Prentice-Wilcoxon test) in \textsf{M0} and is consistent with the previous analysis of the WHI observational data \citep{prentice2005combined}. Compared to their counterparts in \textsf{M0}, treated and matched comparison groups in \textsf{M1} have lower survival probabilities; however, a qualitatively similar cardioprotective effect persists in the design \textsf{M1} ($P$-value = $0.003$). Recall that the design \textsf{M1} has largely mimicked the cardiovascular risk profile of the trial population; therefore, it seems that the difference in cardiovascular risk profile between observational study and trial populations is \emph{not} sufficient in explaining the inconsistency in the trial and observational study results. Compared to \textsf{M0} and \textsf{M1}, the matched design \textsf{M2} best resembles the trial population in their HRT initiation age and previous HRT use; when we examine the survival outcomes in \textsf{M2}, the cardioproctective effect seemed to disappear ($P$-value = $0.495$). 

\textcolor{black}{Our results suggested that once we maximally restricted ourselves to the observational study cohort resembling the RCT population, the comparison between the treated and matched control groups seemed to start aligning reasonably well with that in the RCT. Unfortunately, this finding still cannot lend full credibility to the WHI observational study because of unmeasured confounding heterogeneity: it is conceivable that little unmeasured confounding is present in the 3,000 matched pairs in design $\textsf{M2}$, but sizable unmeasured confounding could persist in the rest of the observational study and still nullify the observational study result. With this important caveat in mind, three comparisons facilitated by three different study designs \textsf{M0}, \textsf{M1} and \textsf{M2} seemed to support \citeauthor{willett2006re}'s \citeyearpar{willett2006re} assessment that HRT initiation age and previous HRT usage played a key role in the discrepancy between WHI observational study and trial findings, and resonate with similar findings in \citet{hernan2008observational}.}

\begin{table}[ht]
\centering
\caption{\small Balance table of baseline covariates in WHI observational study before matching, two matched samples constructed using the proposed algorithm, and the matched template. Match $\textsf{M1}$ constructed $10000$ matched pairs and Match $\textsf{M2}$ $3000$ matched pairs.}
\label{tbl: matched samples I}
\resizebox{\textwidth}{!}{
\begin{tabular}{lccccccccc}
  \hline
  &\multirow{4}{*}{\begin{tabular}{c}\textbf{OBS} \\ \textbf{Treated}\end{tabular}} 
  &\multirow{4}{*}{\begin{tabular}{c}\textbf{Matched} \\ \textbf{Treated} \\ \textbf{M1}\end{tabular}}
  &\multirow{4}{*}{\begin{tabular}{c}\textbf{Matched} \\ \textbf{Treated} \\\textbf{M2}\end{tabular}}
  &\multirow{4}{*}{\begin{tabular}{c}\textbf{Template}\end{tabular}}
   &\multirow{4}{*}{\begin{tabular}{c}\textbf{Matched} \\ \textbf{Control} \\\textbf{M2}\end{tabular}}
   &\multirow{4}{*}{\begin{tabular}{c}\textbf{Matched} \\ \textbf{Control} \\\textbf{M1}\end{tabular}}
    &\multirow{4}{*}{\begin{tabular}{c}\textbf{OBS} \\ \textbf{Control}\end{tabular}}
   &\multirow{4}{*}{\begin{tabular}{c}\textbf{SMD} \\ \textbf{M1}\end{tabular}}
    &\multirow{4}{*}{\begin{tabular}{c}\textbf{SMD} \\ \textbf{M2}\end{tabular}}
   \\ \\ \\ \\ \\ [-0.6em]
   
      \multicolumn{10}{c}{\large\textbf{Sample Size}}\\  \\ [-0.6em]
   \hspace{0.2cm}\textbf{N} &18340 & 10000 &3000 &1000 &3000 &10000 &75303 & \\ \\ [-0.6em]
   
   \multicolumn{10}{c}{\large\textbf{HRT Prior Usage}}\\  \\ [-0.6em]
   \textbf{Age at initiation} \\
   \hspace{0.2cm}NA & 0 &0 &0 &0 &0.88 &0.86 &0.89\\
   \hspace{0.2cm}Age & 53.87 &56.83 &60.67 &61.93 &53.14 &52.44 &52.04\\
   \textbf{Previous E+P use, yrs} &7.02 &5.54 &3.24 &0.70 &0.45 &0.51  &0.47\\ \\ [-0.6em]
   
   \multicolumn{10}{c}{\large\textbf{Covariates Collected in RCT and OBS}}\\  \\ [-0.6em]
\textbf{Age at screening} & 60.84 & 62.21 & 63.57 & 63.10 & 63.35 & 62.15 & 64.30 & 0.01 & 0.03 \\ 

\textbf{Race/Ethnicity}\\
 \hspace{0.2cm}White & 0.89 & 0.85 & 0.84 & 0.84 & 0.84 & 0.85 & 0.82 & -0.00 & -0.01 \\ 
   \hspace{0.2cm}Black/Hispanic & 0.06 & 0.10 & 0.12 & 0.12 & 0.12 & 0.10 & 0.14 & 0.00 & 0.01 \\
   \textbf{Education}\\
 \hspace{0.2cm}College or above & 0.55 & 0.41 & 0.36 & 0.38 & 0.36 & 0.40 & 0.40 & 0.01 & 0.00 \\
  \hspace{0.2cm} Some college & 0.32 & 0.41 & 0.43 & 0.40 & 0.43 & 0.41 & 0.37 & -0.00 & 0.01 \\
  \hspace{0.2cm} High school diploma/GED & 0.11 & 0.15 & 0.17 & 0.17 & 0.17 & 0.16 & 0.17 & -0.01 & -0.02 \\ 
  \textbf{Blood pressure}\\
  \hspace{0.2cm}Systolic & 123.27 & 125.26 & 126.69 & 126.99 & 126.28 & 125.11 & 127.86 & 0.01 & 0.02 \\ 
  \hspace{0.2cm}Diastolic & 74.03 & 74.42 & 75.01 & 75.66 & 74.81 & 74.36 & 74.91 & 0.01 & 0.02 \\ 
  \textbf{BMI} & 25.85 & 27.01 & 28.22 & 28.46 & 27.61 & 26.78 & 27.61 & 0.04 & 0.11 \\ 
 \textbf{Smoking}\\
 \hspace{0.2cm}Current smoker & 0.05 & 0.09 & 0.11 & 0.10 & 0.11 & 0.09 & 0.06 & 0.00 & 0.01 \\  
  \hspace{0.2cm}Never smoked & 0.47 & 0.47 & 0.49 & 0.51 & 0.47 & 0.47 & 0.51 & 0.00 & 0.04 \\
  \hspace{0.2cm}Past smoker & 0.46 & 0.43 & 0.39 & 0.37 & 0.41 & 0.43 & 0.41 & -0.00 & -0.05 \\ 
  \textbf{No. of PA episodes}\\
  \hspace{0.2cm}Total & 5.80 & 5.23 & 4.79 & 4.71 & 4.83 & 5.22 & 5.29 & 0.00 & -0.01 \\
  \hspace{0.2cm}Medium to strenuous & 3.58 & 3.05 & 2.59 & 2.56 & 2.67 & 3.01 & 2.94 & 0.01 & -0.02 \\   
  \textbf{Unopposed estrogen use ever} & 0.12 & 0.10 & 0.10 & 0.10 & 0.10 & 0.10 & 0.43 & 0.00 & -0.00 \\
   \textbf{No ovary removed} & 0.92 & 0.95 & 0.95 & 0.95 & 0.95 & 0.95 & 0.64 & -0.00 & 0.00 \\ 
  \\
   \multicolumn{10}{c}{\large\textbf{Additional OBS Covariates}}\\  \\ [-0.6em]
  \textbf{Region}\\
  \hspace{0.2cm}Midwest & 0.22 & 0.22 & 0.22 &  & 0.24 & 0.24 & 0.22 & -0.04 & -0.04 \\ 
  \hspace{0.2cm}Northeast  & 0.18 & 0.20 & 0.21 &  & 0.24 & 0.21 & 0.24 & -0.03 & -0.07 \\
  \hspace{0.2cm}South & 0.25 & 0.25 & 0.25 &  & 0.24 & 0.24 & 0.26 & 0.01 & 0.02 \\
     \textbf{Partner's education}\\
  \hspace{0.2cm}College or above & 0.43 & 0.37 & 0.32 &  & 0.31 & 0.35 & 0.30 & 0.03 & 0.03 \\ 
  \hspace{0.2cm}Some college & 0.16 & 0.18 & 0.18 &  & 0.18 & 0.18 & 0.17 & 0.00 & -0.01 \\  
   \hspace{0.2cm}High school diploma/GED & 0.07 & 0.08 & 0.09 &  & 0.09 & 0.09 & 0.09 & -0.02 & -0.01 \\ 
   \textbf{Income}\\
   \hspace{0.2cm}Below 35K & 0.23 & 0.29 & 0.35 &  & 0.37 & 0.30 & 0.40 & -0.03 & -0.03 \\ 
   \hspace{0.2cm} 35K - 75K & 0.42 & 0.43 & 0.41 &  & 0.39 & 0.43 & 0.36 & 0.00 & 0.03 \\ 
   \hspace{0.2cm}Above 75K & 0.29 & 0.22 & 0.18 &  & 0.16 & 0.20 & 0.16 & 0.04 & 0.04 \\
   \textbf{Marital status}\\
   \hspace{0.2cm}Married & 0.69 & 0.67 & 0.62 &  & 0.62 & 0.66 & 0.60 & 0.02 & 0.02 \\ 
   \hspace{0.2cm}Divorced/Widowed & 0.26 & 0.29 & 0.33 &  & 0.33 & 0.30 & 0.35 & -0.01 & 0.00 \\
 \textbf{Employment status}\\
  \hspace{0.2cm}Yes & 0.45 & 0.40 & 0.36 &  & 0.35 & 0.40 & 0.32 & 0.00 & 0.01 \\ 
  \hspace{0.2cm}No & 0.53 & 0.58 & 0.62 &  & 0.63 & 0.58 & 0.65 & -0.00 & -0.01 \\ 
\textbf{Reproductive history}\\  
\hspace{0.2cm}Oral contraceptive use ever& 0.53 & 0.48 & 0.43 &  & 0.42 & 0.47 & 0.37 & 0.00 & 0.01 \\ 
  \hspace{0.2cm}OC duration in years & 5.59 & 5.55 & 5.47 &  & 5.18 & 5.32 & 5.16 & 0.06 & 0.08 \\  
  \textbf{Preexisting Conditions}\\
  \hspace{0.2cm}Stroke & 0.01 & 0.01 & 0.01 &  & 0.01 & 0.01 & 0.02 & 0.01 & -0.00 \\ 
  \hspace{0.2cm}MI & 0.01 & 0.02 & 0.02 &  & 0.02 & 0.02 & 0.03 & -0.01 & 0.01 \\ 
  \hspace{0.2cm}CHF & 0.00 & 0.01 & 0.01 &  & 0.01 & 0.00 & 0.01 & 0.02 & 0.01 \\ 
  \hspace{0.2cm}Liver diseases & 0.02 & 0.02 & 0.02 &  & 0.02 & 0.02 & 0.02 & -0.01 & 0.03 \\ 
  \hspace{0.2cm}Hypertension & 0.25 & 0.29 & 0.34 &  & 0.31 & 0.28 & 0.35 & 0.02 & 0.07 \\ 
  \hspace{0.2cm}Fracture & 0.10 & 0.12 & 0.16 &  & 0.13 & 0.12 & 0.14 & 0.03 & 0.09 \\ 
  \hspace{0.2cm}CABG/PTCA & 0.01 & 0.01 & 0.02 &  & 0.01 & 0.02 & 0.02 & -0.01 & 0.04 \\ 
  \hspace{0.2cm}BRCA & 0.01 & 0.01 & 0.01 &  & 0.01 & 0.01 & 0.07 & 0.00 & 0.00 \\
   \hline
\end{tabular}}
\end{table}

\section{Summary}
We proposed a statistical matching algorithm that constructs well-matched pairs from large observational databases that resemble a target population. By designing a representative matched sample, empirical researchers could potentially (i) better reconcile the sometimes conflicting study findings, and (ii) answer a clinical/epidemiological query for a scientifically meaningful population. We applied the proposed method to investigate the discrepancy between the Women's Health Initiative (WHI) observational study and clinical trial findings on HRT's effect on coronary heart diseases. The method facilitated some interesting findings. In particular, we found that a matched cohort study constructed from the WHI observational data still supported a cardioprotective effect of HRT, and this cardioprotective effect persisted even after the designed matched samples were forced to resemble the WHI trial population in their cardiovascular risk profile. However, in a matched design that further resembles the WHI trial intervention group in the HRT initiation age, the cardioprotctive effect seemed to disappear. Our findings provide some evidence for the argument that HRT initiation age might have played an important role in explaining the observational study and trial discrepancy \citep{willett2006re, hernan2008observational}.

\section*{Data Availability Statement}
The data that support the findings of this paper are available from the National Heart, Lung, and Blood Institute. Restrictions apply to the availability of these data, which were used under license for this paper. Data are available from the author with the permission of the NHLBI.

\clearpage
\bibliographystyle{apalike}
\bibliography{match2C}

\end{document}